# Relation between positional & strength asymmetries of double radio sources associated with active galaxies

[**Brief**: Arm length & flux asymmetries for doubles in AGs]

[**Briefer**: Bilateral arm & flux asymmetries for doubles]

*by*

**Dilip G Banhatti**
[dilip.g.banhatti@gmail.com, banhatti@uni-muenster.de]
School of Physics
Madurai Kamaraj University
Madurai 625021
India

## Abstract / Summary
We bring out the identity between two ways of defining a single parameter to combine positional & strength asymmetries of extended extragalactic double radio sources associated with active galaxies. Thus, **$(r.s - 1)/[(1 + r).(1 + s)]$**, combining arm ratio r (defined to be $\leq 1$, i.e., shorter to longer arm) & strength ratio s (in the sense closer to farther, so that it may be <, > or = 1), **is identical to $-(1/2)[(1 - f_r)/(1 + f_r) - \theta]$**, where $f_r$ is strength ratio defined $\geq 1$ (i.e., stronger to weaker), & $\theta = \pm (Q - 1)/(Q + 1)$, $\pm$ signs applying respectively to doubles with closer hotspot fainter & those with closer hotspot brighter, while Q is arm ratio defined $\geq 1$.

## Arm ratios & strength ratios

The extended radio structure disposed around an active galaxy (AG = radio galaxy or quasar) is very often confined to a band on the sky which is many times larger than the optical galaxy size. This band may be almost straight, bent slightly in roughly inversion symmetric S or Z shape, or bent into a C or U shape (e.g., Banhatti 1998 & references therein). For straight & inversion symmetric cases, the radio emission is concentrated towards the edges for the most powerful edge-brightened Fanaroff-Riley (1974) type 2 AGs, while for the less powerful edge-darkened FR1 AGs, it is concentrated toward the inner parts of the band. The positional bilateral (a)symmetry of FR2 AGs is conveniently quantified by arm ratio r defined to be $\leq 1$ (Banhatti 1979, 1980) or $Q \equiv 1/r$, defined to be $\geq 1$ (Teerikorpi 1984, 1986 & references therein). The strength (a)symmetry is measured by the flux density ratio s, in the sense closer hotspot or component to the farther one, so that s may be <, > or = 1. Alternatively, the strength ratio $f_r$ may be defined to be $\geq 1$ (stronger to weaker) (Teerikorpi 1984, 1986).

## Positional offset between AG & radio centroid

If the optical AG position coincides with the centroid for a double, r.s = 1, while r = 1 defines a double with equal arms & s = 1 applies to one with equally strong bilateral components or hotspots. Put another way, r = 1 defines the geometric centre, while r.s = 1 gives the centroid. Straight FR2 AGs' edge-brightened structure leaves little emission outside hotspots. Define by $\delta\varphi$ the displacement from the radio centroid to the optical AG position, so that it is measured positive toward the closer hotspot. Let $S_1, \varphi_1$ & $S_2, \varphi_2$ be the strengths & arm lengths for closer & farther hotspots. Thus $\varphi_1 \leq \varphi_2$, & $(S_1 + S_2).\delta\varphi = S_1.\varphi_1 - S_2.\varphi_2$. Clearly, $r = \varphi_1/\varphi_2 \equiv \varphi_</\varphi_>$, & $s = S_1/S_2$. The total angular size of the straight double is $\varphi = \varphi_1 + \varphi_2$, while total strength $S = S_1 + S_2$. It is easy to see that **$\delta\varphi/\varphi = (r.s - 1)/[(1 + r).(1 + s)]$** (Swarup & Banhatti 1981, Banhatti 1985, 1998). Note that this relation combines arm asymmetry r & strength asymmetry s into a single parameter. The quantity on the LHS of this equality may be estimated for large optically identified powerful radio source samples, even if detailed structures are not known, while the RHS needs knowledge of bilateral structures for evaluation. Swarup & Banhatti (1981) use this with profit to extend their conclusions based on 3CR & Ooty doubles' detailed structures to fainter samples, where only LHS estimates are feasible.

Teerikorpi (1984) defines two types of bilateral symmetry: yes-type satisfying so-called Mackay's (1971) rule that closer hotspot is brighter (CHOB – see Banhatti 1988), & no-type with closer hotspot fainter (CHOF). Further, he defines arm ratio $Q \geq 1$, i.e., $Q = \varphi_2/\varphi_1 \equiv \varphi_>/\varphi_<$ (which = 1/r, see above), ratio of longer to shorter arm of a double, & $\theta = \pm (Q - 1)/(Q + 1)$, with ± signs for CHOF & CHOB doubles. Calling $f_r$ the strength ratio in the sense stronger to weaker, thus $\geq 1$ by definition, $f_r$ & $\theta$ are combined into the single parameter $\xi = -(1/2)[(1 - f_r)/(1 + f_r) - \theta]$. Noting that $S_1 > S_2$ for CHOB doubles, & $S_1 < S_2$ for CHOF doubles, & keeping track of convention for positive $\delta\varphi$, it is easy to show that $\delta\varphi/\varphi$ is identical to $\xi$, i.e.,

**$(r.s - 1)/[(1 + r).(1 + s)] = -(1/2)[(1 - f_r)/(1 + f_r) - \theta]$.**

## Bibliography

Banhatti, D G 1979 *Bull Astr Soc* India **7** 116 … … …
Banhatti, D G 1980 *A&A* **84** 112-4 Expnsn speeds in extdd extragalactic dbl rad srcs fm angular structure
Banhatti, D G 1984/5 PhD thesis TIFR / IIT Bombay Evolution of extragalactic rad srcs
Banhatti, D G 1988 *Astrophys&SpSc* **140** 291-300 Linear size distrbn of dbl rad srcs in The constt speed symmetric expnsn model
Banhatti, D G 1998 *Phys Rep* **303** 81-182 Bilateral symmetry in AGs
Fanaroff, B L & Riley, J M 1974 *MNRAS* **167** 31P- 35P … … …
Mackay, C D 1971 *MNRAS* **154** 209-27 … … …
Swarup, G & Banhatti, D G 1981 *MNRAS* **194** 1025-32 Evoln of extdd extragalctic dbl rad srcs
Teerikorpi, P 1984 *A&A* **132** 179-86 On the asym of dbl rad srcs in qsrs
Teerikorpi, P 1986 *A&A* **164** L11-L12 Rad asym vs size for qsrs with diffrt asym types

[*Readers are welcome to email author with suggestions / comments / queries.*]